\documentclass[twocolumn]{aa}
\usepackage{graphicx}%
\usepackage{txfonts}
\usepackage{natbib}

\begin{document}

\title{Global characteristics of the first IBIS/ISGRI catalogue
  sources: unveiling a murky episode of binary star evolution}

\author{A.~J. Dean \inst{1},  A. Bazzano \inst{2}, A.~B. Hill
  \inst{1}, J.~B. Stephen \inst{3}, L. Bassani \inst{3}, E.~J. Barlow
  \inst{1}, A.~J. Bird \inst{1}, F. Lebrun \inst{4}, V. Sguera
  \inst{1}, S.~E. Shaw  \inst{1,5}, P. Ubertini \inst{2}, R. Walter
  \inst{5} and D.~R. Willis \inst{1}\thanks{\emph{Present address:}
  INTEGRAL Science Data Centre, Chemin d'Ecogia 16, CH-1290 Versoix,
  Switzerland}}

\offprints{A.~J. Dean, email: ajd@astro.soton.ac.uk}

   \institute{School of Physics and Astronomy, University of Southampton,
              Highfield, SO17 1BJ, UK
              \and
              IASF-Rm, INAF, Via Fosso del Cavaliere 100, I-00133 Rome, Italy.
              \and
              IASF-Bo, INAF, Via Gobetti 101, I-40129 Bologna, Italy.
              \and
              CEA-Saclay, DAPNIA/Service d'Astrophysique, F-91191 Gif sur Yvette Cedex, France.             
              \and
              Geneva Observatory, INTEGRAL Science Data Centre, Chemin
              d'Ecogia 16, 1291 Versoix, Switzerland.
   }

   \authorrunning{A.~J. Dean et al.}
   \titlerunning{Global Characteristics of the 1$^{st}$ IBIS/ISGRI catalogue
  Sources} 

   \date{Received 25 May 2005 / Accepted 03 August 2005}

   \abstract{INTEGRAL is the first gamma-ray astronomy mission with a
   sufficient sensitivity and angular resolution combination
   appropriate to the detection and identification of considerable
   numbers of gamma-ray emitting sources. The large field of view ($\sim$
   30$^\circ$ zero response FWHM) enables INTEGRAL to survey the galactic plane on a
   regular ($\sim$weekly) basis as part of the core programme. The first
   source catalogue, based on the 1st year of core programme data ($\sim$5 Msec)
   has been completed and published \citep{ibissurvey}. It contained
   123 $\gamma$-ray sources (24 HMXB, 54 LMXB, 28 ``unknown'', plus 17
   others) - sufficient numbers for a reasonable statistical analysis
   of their global properties. These were located to a positional
   accuracy of typically 0.72 arc minutes. The detection of
   previously unknown $\gamma$-ray emitting sources generally exhibiting high
   intrinsic absorption, which do not have readily identifiable
   counterparts at other wavelengths, is intriguing. The substantial
   fraction (roughly 20\% of the total number) of unclassified $\gamma$-ray
   sources suggests they must constitute a significant family of
   objects. In this paper we review the global characteristics of the
   known galactic sources as well as the unclassified objects with the
   twin aims of investigating how the unclassified set may fit into
   stellar evolution and improving our understanding of known X-ray
   binary systems through the non-thermal $\gamma$-ray channel. In the
   context of the known systems we are very conscious that they
   constitute a $\gamma$-ray selected set, and may exhibit subtle generic
   differences to the rest of the class. We present Log(N)-Log(S)
   distributions, angular distributions, and for systems with reliable
   distance estimates the spatial distributions within the Galaxy and
   luminosity functions. For the unknown sources, this statistical
   analysis has shown that they are most likely to be HMXBs containing
   a highly magnetised neutron star. The lack of X-ray counterparts
   for these sources indicates a high degree of intrinsic obscuration.

   \keywords{Gamma-rays: observations -- X-rays: binaries --  Galaxy:
                              general -- Galaxy: structure --
                              Galaxy: stellar content}
                              }
   \maketitle

%
\section{Introduction and the First IBIS/ISGRI Catalogue}

INTEGRAL is the first gamma-ray astronomy mission with a sufficient
sensitivity and angular resolution combination appropriate to the detection
and identification of considerable numbers of gamma-ray emitting sources
\citep{Winkler}. With an observation time of roughly 1000 seconds
required to detect a 1$M_{\sun}$ neutron star emitting at the Eddington
limit from a distance of 10kpc it is clear that a meaningful survey of
discrete galactic gamma-ray emitting objects is possible in the lifetime of
INTEGRAL. These will be located to a positional accuracy of typically 0.72 arc
minutes. 

A significant fraction of the INTEGRAL core programme is devoted to regular
scans of the galactic plane and a deep exposure of the galactic
centre. This is facilitated by the large field of view
of the on-board telescopes (9$^\circ$ $\times $ 9$^\circ$ fully coded
and 29$^\circ$ $\times $ 29$^\circ$ zero response
for IBIS, \citet{Ubertini}), which permits $\sim$10$^{4}$
second exposures of the galactic plane to be made on a regular basis (%
$\sim$every 12 days). The sample of sources used for this study is
derived directly from the first IBIS/ISGRI catalogue \citep{ibissurvey},
which represents the first year of the INTEGRAL Galactic Plane Survey
(GPS) and Galactic Centre Deep Exposure (GCDE) observations.
The 1$^{st}$ IBIS/ISGRI survey catalogue was compiled based on all Core
Programme observations between revolutions 46 and 120
inclusive. INTEGRAL/IBIS data is organised into short pointings
(science windows) of $\sim$2000 seconds.  OSA
software version 3 was used to create images at science window level,
including two iterations of processing to first produce a cleaning
catalogue, and a second processing to remove all ghosts from the images. The 
$\sim$2500 individual science window images were then mosaiced using a
custom tool to produce deep all-sky maps. An initial source list was
obtained by searching all-sky mosaics constructed in both 20-40 and 30-50
keV bands, using the {\it SExtractor} tool. The 20-40 keV band was optimised for
detection of soft sources, and a threshold of 7 sigma was applied. The 30-50
keV band was judged to produce the cleanest images, and a threshold of 6
sigma was used. By inspection of the image statistics, it was determined
that at maximum, one source in the catalogue would be a false
detection. It has since been discovered that the indentification of IGR
J17460-3047 was in fact a false detection.

Positions for each source were provided using the centroiding function in
{\it SExtractor}, which allows for a non-Gaussian PSF imposed by searching all-sky
maps in the Aitoff projection. The source positions were principally
extracted from the 30-50 keV maps, which were least contaminated by
background structures.  In the case of very soft sources, positions
were taken from the 20-40 keV maps.  Positions for previously known sources were
cross-checked against archive positions, and found to match the expected
point source location accuracy well\citep{Gros}. The typical point source
location error range is 20''--3', with a 1 arc minute error
circle for a source $>$10$\sigma$(90\% confidence).  The precision of
the source locations allow for the clear identification/association to
sources seen by previous missions and hence source classification is
possible.  Additionally, those sources
without previously seen high energy counterparts can be confidently identified as
newly discovered INTEGRAL $\gamma$-ray sources.  Fluxes for each source were
obtained by extraction of values from the all-sky mosaics at the
positions of best fit provided by {\it SExtractor}.

The 1st IBIS/ISGRI survey catalogue is constructed from approximately
5 Msec of observations spread throughout the
period February 2003 to October 2003. The observations were clustered along
the galactic plane; consequently most of the $\gamma$-ray sources are stellar
systems in one form or another. INTEGRAL observes the available galactic
longitude in a sun-orthogonal manner dictating that most regions of the
galactic plane have been observed intermittently during two typically three
month periods within the annual cycle and spaced roughly six months apart.
Since most of the galactic $\gamma$-ray sources are highly variable the
catalogue is a time averaged map and the flux value quoted depends on its
total accumulated flux averaged over the period of the observation for that
region of the sky, no other attempt has been made to correct the recorded
flux to an "average" value. A number of
transient objects are therefore missed if their cumulative emission over the
survey period does not rise above the aggregate threshold. 

Of the 123 sources in the catalogue, 5 were unambiguously identified with
AGN, 5 with white dwarfs, 4 with radio pulsars (free neutron stars), 3
supernovae remnants and one cluster. Of the remainder, 54 were identified
with known low mass X-ray binary (LMXB) systems, 24 with high mass
X-ray binarys (HMXB) and the remaining 28 sources have no firm
classification and are here after described as ``unclassified''.  In this paper we
have only considered the LMXBs, the HMXBs and the unclassified
objects, discounting IGR J17460-3047 as a flase detection; the small
numbers of the remaining object types preclude any meaningful statistical
analysis. Formally, the LMXBs were classified as having the primary star with
a mass $<$ 1 M$_{\sun}$ \citep{Liu_a} and HMXBs having a
primary star with M$>$ 1 M$_{\sun}$ \citep{Liu_b}. Since
the distribution of the unclassified sources is strongly associated
with the galactic structure we have assumed, for the purposes of deriving
LogN-LogS distributions that the contamination of the sample by
extragalactic sources is extremely small, or even non existent. 

The number of galactic gamma-ray sources in the first IBIS-ISGRI catalogue
are comparable to the early X-ray catalogues e.g. \citet{Warwick} and
are sufficient for a preliminary investigation of their global
characteristics on a statistical basis. It should be noted that the sources
presented in the IBIS-ISGRI catalogue are gamma-ray selected and, although
many of them are well known X-ray objects, it is not surprising that there
are considerable differences to catalogues of X-ray sources of corresponding
depth. This is due to the fact that IBIS-ISGRI detects photons at energies
largely unaffected by photoelectric absorption in universal abundance
matter. Hence objects highly photo absorbed in the classical X-ray band are
clearly visible at gamma-ray energies, a factor confirmed by the significant
fraction of sources previously unclassified. There is clearly a thrust to find
out what these unclassified objects actually are and how they fit into the
overall picture of stellar evolution. There are two main methods to achieve
this end. One is to perform follow-up observations on the individual
gamma-ray sources at other wavelengths in order to identify a counterpart
and thus permit more detailed studies, where possible, over a wider spectral
range. The other approach is to study the sample on a statistical basis and
compare their overall characteristics with other known generic sets of
astronomical objects. There is a strong degree of complementarity in the two
methods; here we investigate the global characteristics of the $\gamma$-ray
sample.

\section{The Log(N)-Log(S) Distributions }

\begin{figure}[tbhp]
        \centering
        \includegraphics[width=0.9\linewidth, clip]{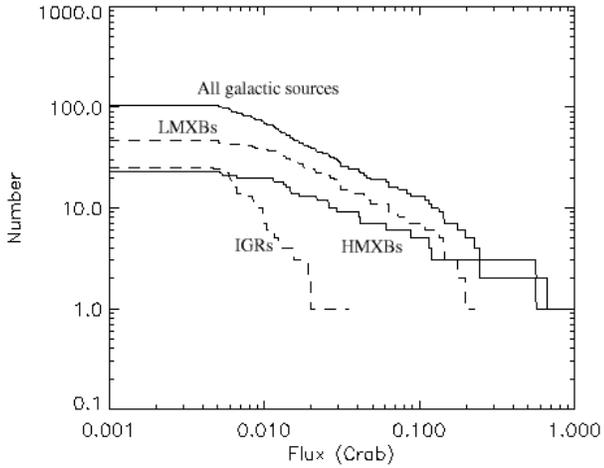}
        \caption{The raw Number-Flux relationship for the sources in
        the 1st INTEGRAL/IBIS survey. The curves shown are for, All
        Galactic Sources; LMXB's, HMXB's and the unclassified sources.}
                \label{Uncorrected}
        \end{figure}

In this section we construct the number flux distribution for the various
source types. We have assumed the usual power-law form of N($>$S)
=KS$^{-\alpha}$ for the relationship. Figure~\ref{Uncorrected} shows
the raw form of these relationships for all galactic sources, the
HMXB, the LMXB and the unclassified sources separately.  The source intensities have been 
converted into 20-40 keV Crab equivalent units.  The
decrease in source density below a few milliCrab is due to the flattening of
the absolute source distribution as the sample is flux limited around that
value. This corresponds to an energy flux of $\sim$3 $\times$ 10$^{-11}$ erg s$%
^{-1}$ cm$^{-2}$ for a Crab-like spectrum.

As the sky coverage of the first
survey is not complete, these curves are only indicative of the true galactic
distributions and must be corrected for both the area of sky covered
and the depth of exposure (or the minimum detectable flux (MDF)) at
each point.  This is not straightforward due to the residual
systematic structures in the sky maps, however it is possible by
inspection of the error and exposure distributions to parameterise the
variation in MDF as a function of exposure, thereby taking into
account the effects of the residual systematic variations.

    \begin{figure}[bhp]
        \centering
        \includegraphics[width=0.9\linewidth, clip]{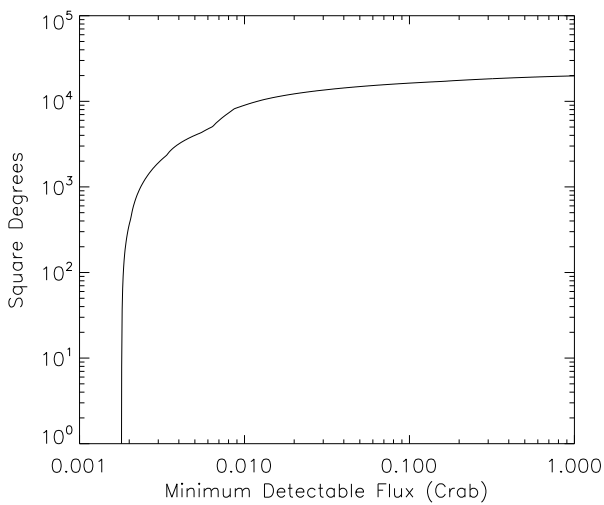}
        \caption{The area of sky observed in the survey as a function
	  of the minimum detectable flux (6sigma).  The flux is in equivalent Crab units.}
                \label{Effarea}
        \end{figure}

    \begin{figure}[bhp]
        \centering
        \includegraphics[width=0.9\linewidth, clip]{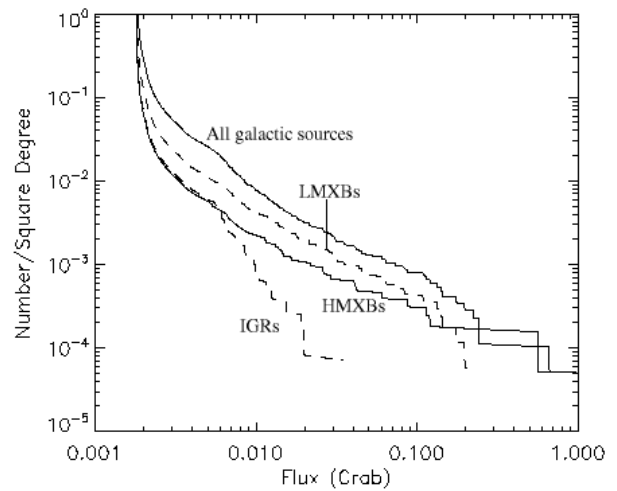}
        \caption{The Number-Flux relationships as shown in Figure 1
        but corrected for exposure and sky area observed.}
                \label{Corrected}
        \end{figure}

From this relationship and the general sky exposure map we can then
construct Figure~\ref{Effarea} which shows the area of sky (in square
degrees) observed as a function of MDF (6 sigma in Crab units). 

We can then use this curve to correct the observed logN-logS resulting
in the relationships shown in Figure~\ref{Corrected}.

The limiting detection threshold of 6 sigma is sufficiently high so that the
maximum likelihood (ML) method \citep{Murdoch} can be
used to calculate the best-fit values of the slope of the
number-flux relationship without the uncertainty in the correction factor
for weaker sources dominating the correction itself. This has been
performed for both the uncorrected and corrected distributions and the
value of the slope found in each case is shown in Table 1.

\begin{table}
\begin{center}
\caption{Slope of Number/Flux relation }
\begin{tabular}{ccc}
\hline
\hline
Source Class & Uncorrected Slope & Corrected Slope\\
\hline
All Galactic Sources & -0.81$\pm$0.10 & -0.91$\pm$0.09 \\ 
LMXB & -0.81$\pm$0.13 & -0.95$\pm$0.13 \\
HMXB & -0.65$\pm$0.15 & -0.81$\pm$0.15 \\
Unclassified Sources & -1.79$\pm$0.37 & -2.11$\pm$0.46 \\ 
\hline
\hline
\end{tabular}
\end{center}
\end{table}

    \begin{figure}[bhp]
        \centering
        \includegraphics[width=0.9\linewidth, clip]{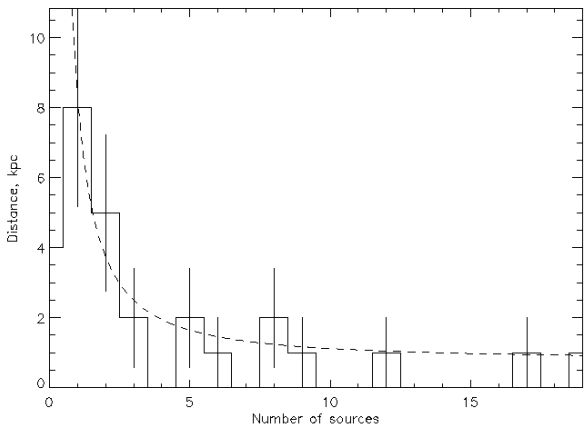}
        \caption{A histogram of the absolute distance of unclassified sources
        away from the galactic centre perpendicular to the line of
        sight (assuming they are $\sim$8 kpc away).  The dashed line
        represents a power-law fit to the histogram.}
                \label{IGR_dist}
        \end{figure}

   \begin{figure*}[bthp]
   \centering
   \includegraphics[width=0.8\linewidth, clip]{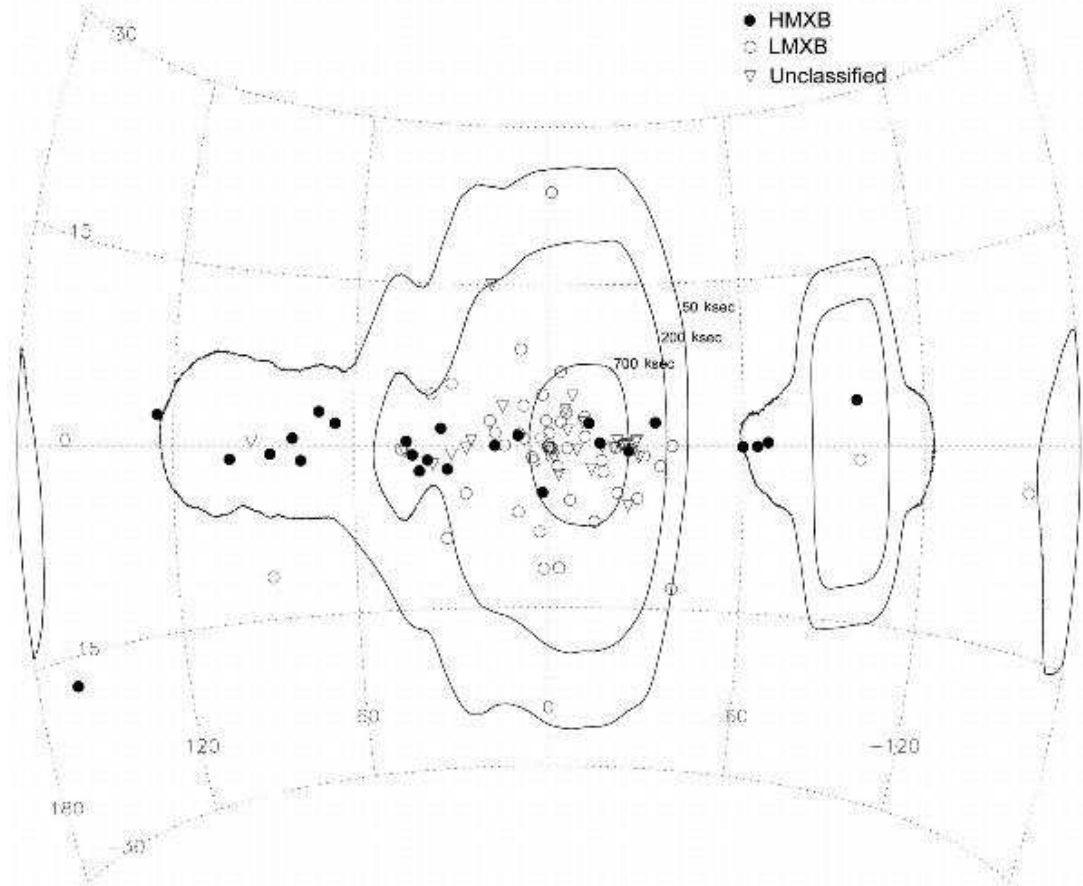}
   \caption{The galactic distribution of sources in the 1$^{st}$
   IBIS/ISGRI catalogue.  Filled circles represent HMXBs; open circles
   represent LMXBs; triangles represent unclassified INTEGRAL
   sources.  Overlaid are contours of exposure time, 50, 200 and 700 ksec.}
              \label{source_dist}%
    \end{figure*}

It is of interest to consider and compare the Log(N)-Log(S)
  distributions of the various sub-groupings. The -0.91 slope of the
  power law for
all the combined galactic sources is close to the -1 value expected for a
uniform infinite plane distribution and lies between the value (-0.79 $\pm$
0.07) measured for galactic sources by ASCA \citep{Sugizaki} in the
classic 2-10 keV X-ray band and the (-1.1) slope derived from the Einstein
galactic plane survey \citep{Hertz}. It is significantly steeper
than the value ($\sim$- 0.5) derived for the bright galactic sources
from the Ariel 5 catalogue \citep{Warwick}. This all-source value is
clearly dominated by the numerically superior LMXBs, which when detached
provide a slope of -0.95 $\pm$ 0.13 as a sub-set. The value for LMXBs
is thus consistent with -1,
as qualitatively expected for a population with a larger scale height
\citep{Sugizaki} and less aligned with the spiral arm tubes.  In
contrast HMXBs are typically located within spiral arms and this would
automatically lead to a value of $\alpha$ closer to 0.5.  In this
context, the -0.81 $\pm$ 0.13 slope measured for the HMXBs appears slightly
flatter than the LMXBs and may reflect the likely location of these objects along
the spiral arms, however the precision of the LMXB and HMXB slopes is
insufficient to distinguish between the two groups. The steep slope
  (-2.11) estimated for the unclassified sources is intriguing
and could imply a different source population at low fluxes, this is far
steeper than we may expect from even an extragalactic contamination.
However, it must be remembered that there are relatively very few of these objects, 
and they are also amongst the weaker sources in the survey therefore care must be taken 
in any interpretation of this slope, as indicated by the large
uncertainty obtained.  Given the angular distribution of the unclassified sample (see
Figure~\ref{longitude}), which indicates a concentration towards the central regions of the
galaxy it could be argued that this steeper distribution may be due to an increase in their 
space density towards the galactic centre. 
        
If we follow this line of reasoning, and if we assume that the newly
discovered INTEGRAL unclassified sources are all located in the galactic 
bulge region we can investigate their spatial distribution by assuming
the distribution is the same in the line of sight towards the galactic
centre as it is orthogonal to our line of sight. We set the zero point
to lie at the peak of the distribution and
take absolute values of the distance of sources from this point because of
the low number of statistics. The resulting distribution is shown in
Figure~\ref{IGR_dist}. The decay in the number of sources away from the Galactic Centre
can be fitted by a power law with an index of $\sim$1.3 $\pm$ 0.8. This
distribution, coupled with the fact that the sensitivity of the first
catalogue does not extend appreciably beyond the galactic centre region,
would imply that for a disk population the sources would obey a LogN-LogS
distribution with a slope of 1.7 $\pm$ 0.4, only about 1 standard
deviation away from the observed value quoted above. 
In general due to the limited number of sources in the first
catalogue, and also the remaining systematics 
affecting the imaging process these results are necessarily subject to
several sources of error which will be overcome with the release of
the second catalogue. Nevertheless they provide qualitative
information on the behavious of galactic source populations above 20 keV.

\section{The Angular Distributions }

The distribution of the $\gamma$-ray emitting sources on the sky is shown in
Figure~\ref{source_dist}. The open circles correspond to the positions of LMXBs, the filled
circles to HMXBs, the unclassified sources are represented by the triangles.
Superimposed on this sky map are contours of the sky exposure. It is readily
apparent that the first year of INTEGRAL observations have been strongly
biased towards the study of the Galactic Plane and in particular the
Galactic Centre, hence the first IBIS/ISGRI catalogue is essentially a
register of galactic sources. Figure~\ref{source_dist} also documents the lack of
uniformity of the exposure along the Galactic Plane. Despite this
non-uniformity, the concentration of the $\gamma$-ray selected HMXBs along the
Galactic Plane is clearly apparent, and likewise the grouping of LMXBs in
the galactic Bulge region.

        \begin{figure}[tbhp]
        \centering
        \includegraphics[width=0.9\linewidth, clip]{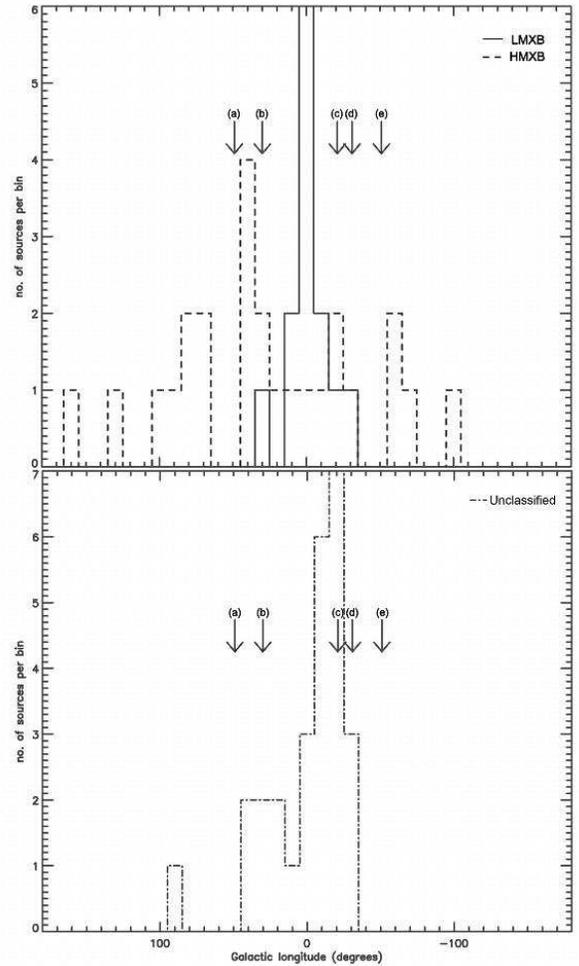}
        \caption{The Galactic longitude distribution of HMXBs, LMXBs
        and unclassified sources.  Labelled are the locations of the spiral arm
        tangents: (a) - Sagittarius; (b) - Scutum; (c) - 3-kpc; (d) -
        Norma; (e) - Centaurus.}
                \label{longitude}
        \end{figure}

Figure~\ref{longitude} reveals the considerable differences in the longitudinal
distribution of the HMXB and LMXB sources. Whereas the LMXB are concentrated
in the Galactic Centre region, the HMXB are spread more extensively along
the Galactic Plane, but in a non-uniform manner, with some evidence for
bunching in the vicinity of the tangential points of the spiral arm
structure. The locations of the spiral arm tangents were taken from
\citet{Englmaier}. Additionally, a number of HMXBs are seen at {\it l$^{II}$} 
$\sim$ $\pm$ 90$^o$, which is indicative of systems located within
our particular spiral arm. This association had been previously noted
through inspection of the Ginga data by \citet{Koyama} and in the
RXTE/ASM data by \citet{Grimm}. 

The association with the spiral arms is entirely to be expected since the
high mass binaries are young stellar systems and should be attached to
regions where star formation has recently taken place, such as the spiral
arms. We will come back to this point later, when we discuss the spatial
distribution of the $\gamma$-ray HMXBs.

Upon first inspection the unclassified INTEGRAL sources appear to have a
longitude distribution concentrated around the Galactic Centre, similar to
that of the LMXBs. To some extent this is misleading as the Galactic Centre
has had the most sky exposure and hence we are more likely to detect new
systems in this region. However upon closer examination, whilst some emulate
the distribution of LMXB, it can be seen that the unclassified sources do not have a
symmetrical distribution about the Galactic Centre and show some tendency to
cluster around the locations of spiral arm tangents, specifically the
Scutum, 3-kpc and Norma arms, as shown in
Figure~\ref{longitude}. Specifically the unclassified sources appear
to concentrate in the 3-kpc and Norma spiral arms. 

        \begin{figure}[tbhp]
        \centering
        \includegraphics[width=0.95\linewidth, clip]{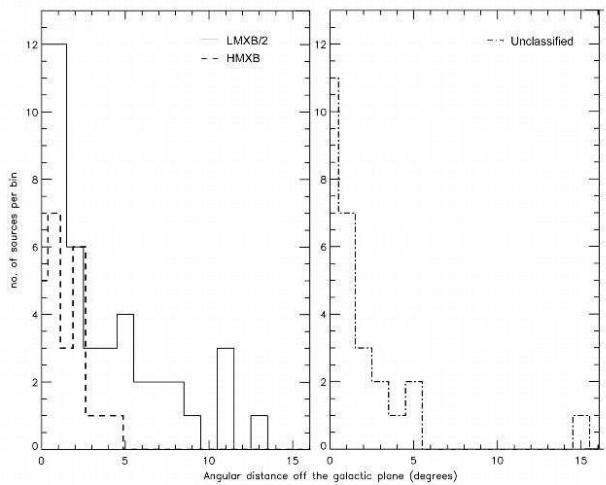}
        \caption{Angular distribution of sources away from the Galactic
        Plane.  Note that the number of LMXBs is divided by 2.}
                \label{latitude}
        \end{figure}

   \begin{figure*}[bthp]
   \centering
   \includegraphics[width=13cm, height=13cm, clip]{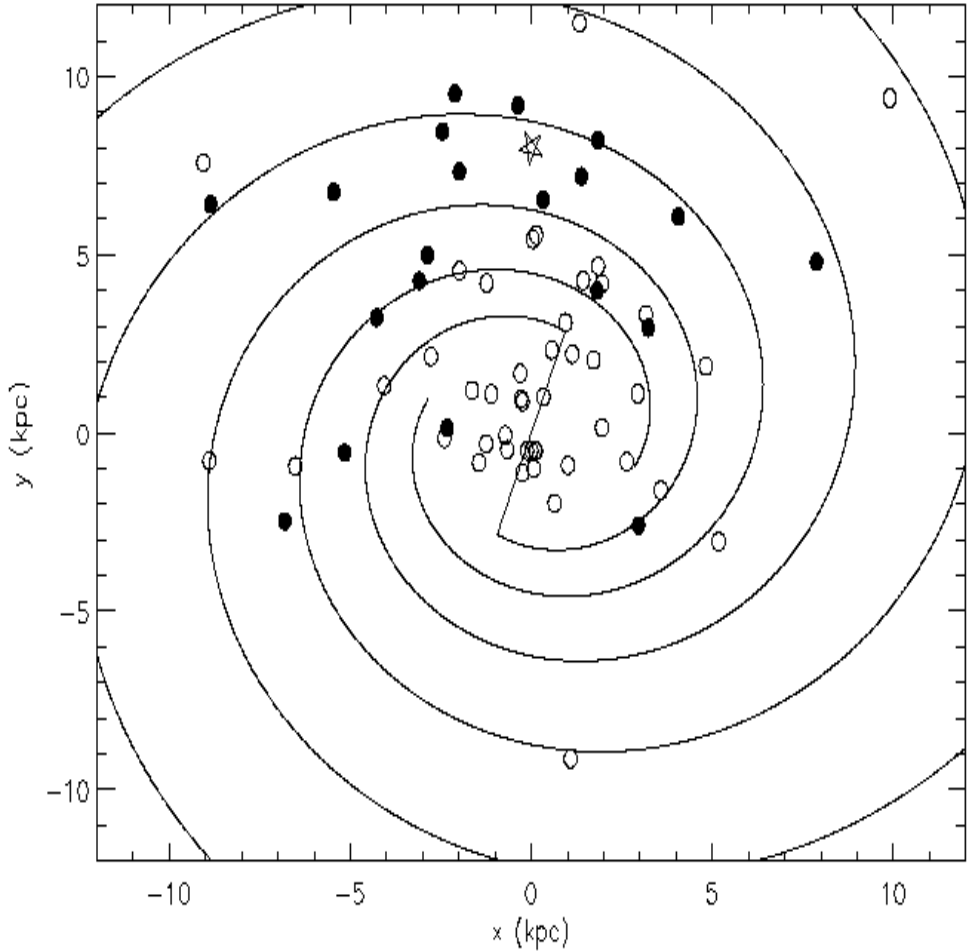}
   \caption{The galactic distribution of the subset of HMXBs (filled circles) and
   LMXBs (open circles) for which distance estimates are available in
   the literature. The Sun is represented by the star symbol and
   is 8 kpc away from the galactic centre.  Superimposed is the 4-arm
   sprial model of \citet{Vallee}.  Source distance errors are on
   average of the order of a kpc.}
              \label{galaxy}%
    \end{figure*}

The angular distribution off the Galactic Plane of the sources is seen in
Figure~\ref{latitude}. The HMXBs exhibit a much narrower range of angular separations as
expected if they are confined to the Galactic Disc. The much wider spread
seen in the LMXBs is indicative of a population derived from the Galactic
Bulge. Predictably the higher spread in the latitude distribution of the
LMXB as opposed to the accumulation of HMXB closer to the Galactic Plane is
to be expected from the relative ages of their progenitor stars. The unclassified
sources appear to have a sharp latitude cut-off similar to that of the
HMXBs, this may indicate that they too are a primarily a population from the
Galactic Disc. The unclassified sources, however, do not precisely conform to either
of the subgroups; this is not unexpected as they could consist of a mixture
of both, or be a separated generic set. Although the longitude distribution is not
entirely dissimilar to the LMXBs, this is probably
distorted by an effect of exposure and the source concentration around
spiral arm tangents suggests an association with high mass
systems. This association is further supported by their latitude profile that more
closely resembles that of the HMXB. However the unclassified sources are very much
associated with the Galactic Centre region, this association may be generic
in some way and possibly an artefact of the selection of the sources through
the $\gamma$-ray channel, which is not sensitive to the effects of higher
photoabsorption generally suffered through the conventional X-ray
observations.

\section{The Spatial Distributions }

The distances to a number of the HMXBs and LMXBs were found through an
examination of the available data as published in the literature
\citep{White, Macomb, Kuulkers}, thus allowing the construction of a 3D model of the
source distribution. An updated list of distances collected from available
literature for neutron stars and black holes in LMXB as reported by
\citet{Jonker} represented one key source of distance information.
Different methods were used in deriving the distances of such objects. For
neutron stars characterized by the presence of type I bursts, distance can
be inferred from observations of Eddington limited bursts. Type I bursts in
Low Mass X-ray Binaries are thermonuclear flashes on the neutron star
surface \citep{Lewin_a}. During some type I bursts the energy
release is high enough such that the luminosity at the surface of the
neutron star reaches the Eddington limit (for a typical neutron star the
Eddington luminosity value ranges between 2 -- 4 $\times$ 10$^{38}$
ergs cm$^{-2}$ s$^{-1}$
). During these events the neutron star's atmosphere expands because of
radiation pressure. In this expansion and subsequent contraction the
luminosity stay constant near the limit and assuming a pure black body
radiation there is a drop in temperature during the expansion of photosphere
radius. The enormous number of bursts and new burst sources discovered with
the WFCs/BeppoSAX gave a very powerful tool in this respect (see
e.g. \citet{Cocchi, Kuulkers, Zand}). In other cases the spectral properties of
the companion star have been used. Where possible \citet{Jonker}
compared the two different methods and the discrepancies seem to
indicate that a larger distance is derived from the first method while using the
second approach the distances are underestimated, possibly indicating
erroneous spectral classification of the companion star due to fast
rotation. For the black hole candidates in our list three different methods
have been used to derive distances, namely by using the interstellar
absorption properties, from the observed proper motions of receding and
approaching lobes when symmetric jets are present, and from the comparison
of the absolute magnitude of the donor star with the apparent magnitude
using radius, spectral type and luminosity as determined directly from
observational data. 

A model of the galaxy taken from \citet{Vallee} is shown in Figure~\ref{galaxy} with the
location of HMXB (filled circles) and LMXB (open circles). It is immediately
obvious that the HMXBs and LMXBs do not share the same spatial
distribution.  The LMXBs can be seen to cluster at the Galactic Centre
as implied by their angular distribution discussed in the previous section.
The LMXBs are clearly clustered in the central $\sim$3 kpc of the
galaxy, while the HMXBs are found outside that radius, in the region
associated with the spiral structure.  However, for those systems $>$3
kpc from the Galactic Centre, it is impossible to say whether LMXB or
HMXB follow the spiral structure more strongly. Theoretically the
appearance of the HMXB epoch should lag behind the leading edge of spiral
arms. This is brought about by the density wave inducing star formation and
the HMXB phase subsequently developing some time later. A broad time window is to
be expected since the period of elapsed time from star formation to the
appearance of a HMXB system is dependent upon the initial masses of its
components, and would naturally reflect this spread. For example
\citet{Tauris} calculate an age of 24.6 million years for a system
which starts with 14.4 M$_{\sun}$ and 8 M$_{\sun}$ stars, and
\citet{van_den_Heuvel_a} calculate an age of 15 million years for a system which
starts with a 16 M$_{\sun}$ and 3 M$_{\sun}$ star.

An investigation of this systematic time lag is beyond the scope of
this paper.  An analysis of this would require a larger population of
HMXBs with well defined distance measures and a model of the galaxy
which incorporates the quasi-stationary spiral structure hypothesis
proposed by \citet{Lin_b}. The key parameter in modelling the spiral
structure is $\Omega _{P}$ the angular velocity of the spiral pattern.
Recent measurements of $\Omega _{P}$ based upon observations of open
clusters by \citet{Dias} strongly indicate that the spiral pattern
rotates as a rigid body with $\Omega _{P}$ $\sim$25 km s$^{-1}$ kpc$%
^{-1}$at the distance of the Sun from the galactic centre. 

\begin{figure}[bhtp]
        \centering
        \includegraphics[width=0.85\linewidth, clip]{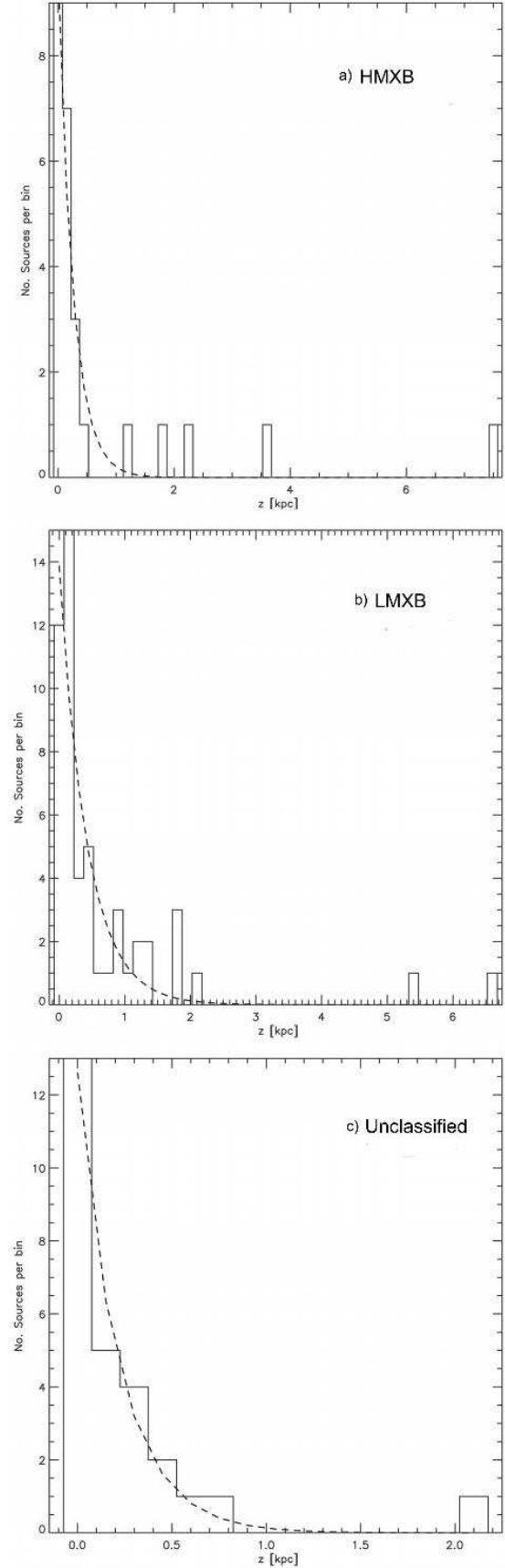}
        \caption{The distribution of the height off the plane of: a)
        HMXBs; b) LMXBs; c) Unclassified sources.  The dashed lines represent
        exponential fits.}
                \label{scale_heights}
        \end{figure}

The scale heights of the HMXB and LMXB systems are shown in Figure~\ref{scale_heights}a and
Figure~\ref{scale_heights}b. In each case a simple exponential decrease in numbers as a
function of the absolute distance h "above" the galactic plane, N = k e$^{-\alpha h}$,
was used to model the data sets. Making the simplistic assumption that all
the unclassified sources reside close to the galactic bulge region, we constructed a pseudo
scale height distribution for these unclassified objects as shown in Figure~\ref{scale_heights}c.
Table~\ref{tab:scales} shows the associated values for the scale height distributions of
the three catagories.

   \begin{table}[tbhp]
   \begin{center}
     \caption{Summary of the chracteristics of the principal
     populations of the IBIS-ISGRI sources in terms of their
     distributions above the Galactic Plane.}
    \begin{tabular}[h]{|l|ccc|} \hline
                       &  k & $\alpha$ & Scale Height (pc) \\
      \hline
      HMXBs             &   9 $\pm$ 3  & 4 $\pm$ 1   & 240$^{+90}_{-40}$ \\
      LMXBs             &   13 $\pm$ 3  & 2.9 $\pm$ 0.6 & 350$^{+90}_{-60}$ \\
      Unclassified sources              &   12 $\pm$ 3  & 4 $\pm$ 1   & 240$^{+90}_{-40}$ \\
      \hline
      \end{tabular}
     \label{tab:scales}
     \end{center}
   \end{table}

It is anticipated that the scale heights of the HMXB and LMXB
  populations are different, and that the vertical distributions
  reflect the age disparity of their parent stars.  Unfortunately, the
  precision of the observed scale heights shown in
  Table~\ref{tab:scales} is such that the HMXB, LMXB and unclassified
  sources have scale heights which are within one sigma of each other
  and are hence indistinguishable.
However it is interesting to note that the scale height of this $\gamma$-ray
selected HMXB set is consistent with the value of 150pc found for the
HMXB selected from RXTE/ASM data \citep{Grimm}. The lack of errors on
the HMXB scale height as measured by \citet{Grimm} make it difficult to
accurately compare the values. Likewise the vertical
distribution of the $\gamma$-ray selected LMXBs is compatible with
the 410$^{+100}_{-80}$ pc scale height of the RXTE/ASM sample of the
LMXB disk population however, our HMXB scale height lies only
  $\sim$2$\sigma$ from this measurement as well. If we take the unclassified source "scale
height" at face value, then it is very interesting to note
that, whilst they appear to conglomerate in the galactic bulge region, their
vertical distribution is the same as the HMXB systems, reflecting the
similarities found in their longitude distributions.

\section{Luminosity Functions }

Many the sources in the first IBIS/ISGRI catalogue have good distance
measurements and it is possible to construct their associated luminosity
functions. However corrections are required to account for non-uniform
exposure, the flux limited nature of the sample and any incompleteness in
identifying object classes. This correction is performed by constructing a
model of the stellar mass distribution of the galaxy as performed by
\citep{Grimm}. A galaxy model consisting of a disk and bulge component was
constructed using the bulge and disk models of \citet{Dehnen}:

\begin{equation}
  \rho_{Bulge} = \rho_{0, Bulge} \cdot \left(
  \frac{\sqrt{r^{2} + \frac{z^2}{q^2}}}{r_{0}} \right)^{-\gamma} \cdot
  exp \left( -\frac{r^{2}+ \frac{z^2}{q^2}}{r_{t}^{2}}\right)
\label{bulge_density}
\end{equation}

\begin{equation}
  \rho_{Disk} = \rho_{0, Disk} \cdot 
  exp \left( -\frac{r_{m}}{r} - \frac{r}{r_{d}} - \frac{|z|}{r_{z}}\right)
\label{disk_density}
\end{equation}

Where $\rho_{0,Bulge}$ and $\rho_{0,Disk}$ are the normalisations, r
is the distance in the plane to the galactic centre, z is the distance out
of the galactic plane and R is the distance from the galactic centre in
spherical coordinates. The other parameters are: q is the oblateness of the
bulge; r$_{0}$ is the scale length of the bulge; r$_{t}$ is the truncation
radius of the bulge; r$_{d}$ is the scale length of the disk; r$_{z}$ is the
vertical scale of the disk; r$_{m}$ is the inner disk cut-off. These
parameters were taken from fit of RXTE/ASM XRBs to the galactic model by
\citet{Grimm}. However we adjusted r$_{z}$ to match our measured scale
heights. All distances are in kiloparsecs. The standard galaxy model uses a
mass ratio of 2:1 for disk:bulge. To correct the LMXB population both model
components were used whereas the HMXB population uses only the disk
component.

To correct the observed luminosity function we calculate the fraction of the
galactic mass which is observable at a given luminosity. The correction is
taken from \citet{Grimm} and is shown below:

\begin{equation}
  \frac{dN}{dL} = \left( \frac{dN}{dL} \right)_{obs} \times \frac{M_{tot}}{M(<D(L))} 
\label{lumin_cor}
\end{equation}

Where $\frac{dN}{dL}$ is the true luminosity function,
$\left(\frac{dN}{dL}\right)_{obs}$ is the observed
luminosity function, $M_{tot}$ is the total mass of the galaxy, M($<$%
D) is the mass of the galaxy inside a distance D from the Sun. D(L) is
defined as:

\begin{equation}
  D(L)= min\left( \sqrt{\frac{L}{4 \pi F_{lim}}}, D_{max} \right)
\label{lumin_dist}
\end{equation}

        \begin{figure}[tbhp]
        \centering
        \includegraphics[width=0.9\linewidth, clip]{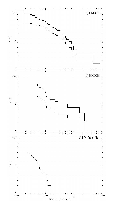}
        \caption{The 20 -- 100 keV luminosity functions of: a) LMXBs;
        b) HMXBs; c) Unclassified sources.  The thin line represents the uncorrected
        luminosty function.  The thick line represents the corrected luminosity
        function as described in the text.  The unclassified sources are
        uncorrected as their source population is undetermined.}
                \label{lumin1}
        \end{figure}

Where F$_{lim}$ is the limiting flux of the sample and D$_{max}$ is the
maximum distance from the Sun of the sources used in constructing the
luminosity function. We take F$_{lim}$ as $\sim$5mCrab in the 20-100
keV energy band which corresponds to $\sim$8.5 $\times $ 10$^{-11}$ ergs cm$
^{-2}$ s$^{-1}$.  The maximum distance at which we detect sources
approximates to the galactic centre distance. As a result this means
that it is only at relatively low luminosities that D$_{max}$ does not
correspond to D(L). As previously discussed the majority of our
exposure is in the vicinity of the galactic centre. To compensate for
the uneven exposure we modified F$_{lim}$ for the four quadrants of
the galactic plane. As the sensitivity is proportional to the square
root of the exposure and that the maximum depth of our sample is 
$\sim$8 kpc the change in the luminosity function correction within
our range of exposures is $<$10\%

        \begin{figure}[tbhp]
        \centering
        \includegraphics[width=0.9\linewidth, clip]{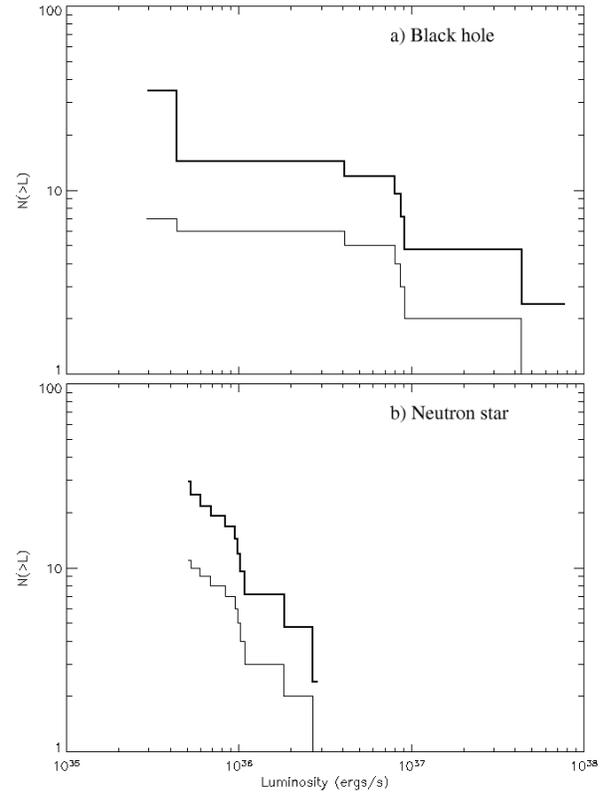}
        \caption{The 20 -- 100 keV luminosity functions of the
        IBIS/ISGRI sources separated according to the nature of the compact object: a) black holes;
        b) neutron stars.  The thin line represents the uncorrected
        luminosty function.  The thick line represents the corrected luminosity
        function as described in the text.}
                \label{lumin2}
        \end{figure}

Accreting X-ray binaries are the brightest class of X-ray sources and their
emission properties are determined by the nature of the compact object i.e.
BHC or neutron star. The strength and geometry of its magnetic field on the
neutron stars are also important as is the geometry of the accretion flow
from the companion. A neutron star with a strong magnetic field ($\sim$%
10$^{12}$ Gauss) will in fact disrupt the accretion flow at several hundred
neutron star radii and material is channelled to the magnetic pole. In
the case
of low magnetic field ($<$ 10$^{10}$ Gauss) the disk may approach close
to the neutron star surface. HMXBs typically have the accreting compact star
as a high magnetic field NS or BH and LMXBs have typically a low magnetic field
NS or BH. Strongly or weakly magnetized neutron stars are identified by the
presence or absence of pulsations in the light curve of the system. X-ray
pulsars are seldom in LMXB, which generally show X-ray bursts that are
suppressed when B is $>$ 10$^{11}$ Gauss \citep{Lewin_b})
and this also explains why bursts are not detected in HMXB systems. For a
recent review on accreting neutron star and black hole properties see
\citet{Psaltis}, which also lists the energies of cyclotron lines in
accretion powered pulsars used to determine the values for the magnetic
field. Magnetic field values for X-ray pulsars in the RXTE/BeppoSAX era are
reported in \citet{Coburn}.

Figure~\ref{lumin1} shows the cumulative $\gamma$-ray (20 -- 100 keV) luminosity
functions for LMXB, HMXB corrected as outlined in the above text. Also
included is a "luminosity function" for
the unclassified sources, assuming that they are all at a common distance of 8
kpc. It is noticeable that, whereas the low mass systems spread their 20 --
100 keV luminosities relatively consistently up to and beyond
10$^{37}$ ergs cm$^{-2}$ s$^{-1}$, the majority of the high mass systems have a strong
inclination to reside around 10$^{36}$ ergs cm$^{-2}$ s$^{-1}$. The
distribution of the unclassified sources exhibits the low luminosity
characteristics of the low mass systems, but also a tendency to favour
luminosities close to 10$^{36}$ ergs cm$^{-2}$ s$^{-1}$. Whereas Figure~\ref{lumin1} presents the 
$\gamma$-ray luminosity functions of the known objects selected by the donor
system, Figure~\ref{lumin2} shows the luminosity functions of known sources grouped by
the nature of the compact object, black hole systems and neutron
stars. There are marked differences in their individual
distributions. Whereas most of the 20 -- 100 keV luminosities of black hole
systems exist around 10$^{37}$ ergs cm$^{-2}$ s$^{-1}$, neutron stars do not emit 
$\gamma$-ray fluxes much in excess of 10$^{36}$ ergs cm$^{-2}$ s$^{-1}$. Although
subdividing the classifications has resulted in dwindling numbers of
examples and hence statistical significance, it is interesting to note
 from Fig.~\ref{lumin1}c and Fig.~\ref{lumin2}, that the unclassified sources are most
likely neutron star systems with a significant fraction, but probably not
all, cohabiting with high mass companions. The use of a $\sim$10$^{37}$
ergs cm$^{-2}$ s$^{-1}$ soft $\gamma$-ray boundary to separate neutron star and black hole
systems is not a new concept. \citet{barret} found
for high luminosity-high energy emitters, the $\sim$100 keV luminosity
of black hole systems is commonly in the range 1 -- 6 $\times $ 10$%
^{37}$ ergs cm$^{-2}$ s$^{-1}$, whereas for neutron star systems it is $\leq $10$^{37}$
ergs cm$^{-2}$ s$^{-1}$, as confirmed by INTEGRAL

\section{Summary and Discussion }

The INTEGRAL galactic plane survey and ensuing first catalogue has offered
for the first time the possibility to investigate the global $\gamma$-ray
characteristics of galactic sources on a reasonable statistical basis. In
this paper we have studied the generalised observational features of $\gamma$%
-ray selected previously unclassified objects, and compared them with the
corresponding parameters of known HMXB and LMXB systems. Although the sample
of sources has been $\gamma$-ray selected the Log(N)-Log(S), angular and
spatial distributions of the known classes, as expected, emulate those of
X-ray selected sets e.g. \citet{Grimm}, there appear to be subtle
differences. Of great interest however are the 27 "unidentified" objects, which naturally have overhanging
questions such as what are they and how do they fit into mainstream stellar
evolution?

Clearly a vigorous programme of follow-up studies on individual objects,
using such missions as XMM-Newton, Chandra, and RXTE for timing
studies, is one means to pursue
the answers to these questions, and such activities are already underway.
\citet{Walter, Hill, Stephen, Revnivtsev, Matt, Rodriguez}, and a number of
identifications with objects that emit at other wavelengths have been made.
The complementary statistical approach employed here essentially comes to
the same conclusion i.e. that a large fraction of the unclassified objects are
obscured high mass X-ray binary systems. Detailed studies of the main
distributional characteristics of the unclassified sources when compared to the
equivalent distributions of known X-ray binaries all point in this
direction. Their Log(N)-Log(S) and broad angular distributions are
consistent with them inhabiting the central 3-4 kpc region of the galaxy,
similar to many of the LMXBs. However a closer look at the angular profile
shows that it is not the central bulge that determines their finer scale
longitude distribution but instead the structure of the inner spiral arms,
indicating a considerable number of the unclassified objects are associated with
younger stellar systems, and hence probably HMXB. Likewise the galactic
latitude distribution of the unclassified sources strongly reflects that of HMXB.
More tenuously, the luminosity function of the unclassified sources, assuming them to
be close to the galactic centre, exhibits features that are similar to HMXB
systems and possibly those incorporating highly magnetised pulsars rather
than black holes as the compact companion. 

It is hardly surprising that INTEGRAL should discover a population of
previously unnoticed sources. The catalogue is $\gamma$-ray selected, and
INTEGRAL operates above the energy threshold for which significant
photoabsorption takes place in universal abundance matter, so that a strong
emitter above $\sim$30keV can be rendered insignificantly weak in the
classical X-ray band. As discussed by \citet{Lutovinov_a}, the HMXB
systems constitute the most likely candidates. The stellar wind accretion
mechanism that dominates in HMXB systems, as opposed to the Roche lobe
overflow associated with LMXBs, naturally provides a suitable dense and
strongly absorbing circumstellar wind to veil the X-ray emission in a
similar manner to the case of the Seyfert 2 AGN configuration \citep{Turner, Malizia}. 

Exactly how the $\gamma$-ray selected highly absorbed systems fit into the
overall picture of binary star evolution is currently unclear.  This
is principally due to the requirement of a suitable companion able to provide
a sufficiently strong stellar wind mass loss rate ($\dot{M}$ $\geq $ 10$^{-4}$ M
$_{\sun}$ yr$^{-1}$) that is capable of generating a suitably
dense gas surrounding the compact object to stifle the X-ray
emission. We need to understand why they are different to "
normal" HMXB. Do these highly absorbed systems relate to the
mass/giant nature of the primary star, or to the orbit configuration, or are
they experiencing a phase the binary systems routinely pass through; but
have not been exposed at other wavelengths? Do they need to be HMXB? Clearly
a series of dedicated observations are required to solve this problem. If
they are not all HMXB systems, then could some of them be intermediate
mass X-ray binaries (IMXBs) \citep{van_den_Heuvel_b}. Some of the
sources could be systems that are passing through,
or close to the common envelope phase \citep{Iben, Taam}. However
statistical grounds this seems unlikely due to
the requirement that their relative numbers need to compatible with the
relatively short timescale ($\leq $ 10$^{3}$ yr) of this phase, which is
dictated by the drag force created as the compact star moves through the
envelope of the extended companion.

Clarification of the above general issues raised by the INTEGRAL source
discoveries will also facilitate a better understanding of the relative
production of neutron stars and black holes. There is uncertainty in the
threshold mass for core collapse into a black hole. Lower mass helium star
systems should also be considered \citep{Brown}. A
fundamental aspect in this context is whether, during the mass loss period,
the helium core remains surrounded by a thick hydrogen mantle, so that
common envelope evolution proceeds \citep{Nelemans}. It
is also possible that a neutron star spiralling inwards in a common envelope
might rapidly accrete and eventually collapse into a black hole
\citep{Chevalier}. On a statistical basis INTEGRAL will provide a clearer picture of the
relative numbers of the various sub species of X-ray binary systems, and
consequently a more balanced understanding of the rather murky final stages
of binary star evolution and by implication eventually more reliable
estimates for the relative numbers of black hole and neutron stars.

\begin{acknowledgements}
      This work is based on observations obtained with INTEGRAL an ESA
      science missions with instruments, science
      data centre and contributions funded by the ESA member states
      with the participation of the Czeck Republic, Poland, Russia and
      the USA.

      The Southampton group acknowledge support through PPARC research
      grants.  A.B.H. acknowledges funding support from a PPARC PhD
      studentship. A.B, L.B., J.B.S. and P.U. acknoweledge the Italian
      Space agency financial support via contract I/R/046/04.
\end{acknowledgements}

\end{document}